# Higher ($2^{nd}$)-order polarization-Wigner function for 'even' entangled bi-modal coherent states

Ravi S. Singh[1#], Sunil P. Singh[2], Lallan Yadava[1#] and Gyaneshwar K. Gupta[1*]

[1]Department of Physics, D.D.U. Gorakhpur University, Gorakhpur-273009 (U.P.) India
[2]Department of Physics, K.N.I.P.S.S., Sultanpur (U.P.) India

Higher ($2^{nd}$)-order Wigner distribution function in quantum phase space for entangled bi-modal coherent states, a representative of higher ($2^{nd}$)-order optical-polarization, is introduced by generalizing kernel (transiting) operator in Cahill-Glauber C(s)-correspondence rule. The nature is analyzed which reveals the occurrence of oscillating three peaks: 'two' for individual bi-modes and third for interference between modes. Also, the graphics of $2^{nd}$-order polarization-Wigner distribution function, incisively, demonstrates that it is of non-Gaussian nature attaining non-negative values in quantum phase space.

**Introduction**

The dynamical state of an incompletely-known quantum system is, in general, specified by density operator (matrix) imbibing all informations pertaining to the system. Stringently, its precise specification is possible only up to a largest complete set of compatible (commutable) dynamical variables. Obviously, this alleged set does not constitute an exhaustive set of all dynamical variables because of intrinsic uncertainties in their measured values dictated by Heisenberg uncertainty principle [1]. But, (quasi)probability distribution function furnishes alternative route to completely characterize a dynamical state of the quantum system in quantum phase space (Planck space). Although any normalized real or complex [2] function in phase space may possess candidacy to serve as probability distribution function, an exquisite distribution function introduced by Wigner in 1932 [3], has nifty edge for the visualization of the dynamics of a quantum system in quantum phase space [4]. Quantum Optics witnessed plethora of probability distribution functions, namely, Glauber-Sudarshan P-function, Husimi Q-function, positive P-function etc. [5]. Moreover, the reconstruction of Wigner function through Radon transformation of marginal distribution function measured experimentally in homodyne detection [6], offers essential suitability over that of Glauber-Sudarshan distribution function of which measurement is recently reported [7]. A great deal of interest has been shown in recent years, due their importance to test quantum non-locality [8], in generalized quasiprobability distribution function discovered by Cahill and Glauber [9] by parameterizing Moyal-Weyl correspondence rule [10]. Worth-mentioning progress has flourished in numerous attempts for constructing Wigner function corresponding to a quantum system in finite dimensional Hilbert space [11], a quintessence for scalable quantum computation [12]. A parentage between Heisenberg- and Weyl-group by means of quantum Fourier transforms has been traced in which a unified approach in the theory of continuous and discrete (finite) quasiprobability distribution function has been chalked out [13].

The Wigner distribution function in signal processing is introduced for the first time by Vilee [14] following seminal works due to Gabor [15] in theory of communication of electronic signals varying randomly in frequency and time. The third-order Wigner-Ville distribution function is derived by Gerr [16] through extension of time-frequency space to time-frequency-frequency space. Latter on several variant of higher-order Wigner functions (HOWF) and their intervening relationships have been carried out by electronics community [17]. Precisely, an avid observation connotes that these HOWF are distribution functions in either mono-time multi-frequency space or multi-time mono-frequency space. Formulation of HOWF in quantum optics has not been attempted although the higher-order moments of Wigner function is being applied in characterizing non-classical traits of optical field [18].

The polarization of light is an earmark of electromagnetic radiation ensuring its transversal nature. Although the concept of polarization is centuries-old, its precise nature still demands ingenious articulation, especially, for the case of optical quantum fields. A recent pellucid study on optical-polarization has exhibited conflicting yet complimentary approaches: operational-measures and computable-measures [19]. The operatic-measure is based upon Stokes parameters (operators) which quantifies states of optical-polarization on the surface of unit-Poincare sphere [20]. Quite earliest [21] as well as studies in last decade [22] have demonstrated that Stokes parameters are insufficient in assessing polarization of light. Also, Stokes theory misleads while characterizing multiphotonic optical quantum field possessing higher-order polarization (Malus law) [23]. Notably, an impeccable criterion is laid down in terms of minimal fluctuations of Stokes parameters on Poincare sphere [24] which, ostensibly, captures only 'special' kind of forth-order Bosonic correlation functions of optical field in which number of creation operators equals to that of annihilation operators [25]. Moreover, the distance-based [26] approach introduces an abstract 'notion of distance' of quantum states of light from that of unpolarized light [21] which specifies variant degrees of polarization for the same quantum state. The inadequacy of Stokes theory and sheer lack of correspondence of distance-based approach to classical theory of optical-polarization spurred a radically different approach on optical-polarization [27, 28] which not only stems from the tenets of classical theory of optical-polarization but also deciphers the twin characteristics optical-polarization parameters of the optical field: 'ratio of real amplitude' and 'difference in phases' specifying a point on the Poincare sphere [29].

Tracing mathematical similarities of Stokes operators with those of Jordan-Schwinger spin angular momenta [30] of two quantum oscillators, a quasi distribution function (SU(2) Q-function) for describing optical-polarization on Poincare sphere is deduced of which 'distance' from the uniform distribution of unpolarized light is ascribed to define a degree of polarization in Quantum Optics [31]. Since it has, vociferously, been pointed out [32] that Stokes operators describing states of optical-polarization found dissimilar Hilbert space for its operation unlike to those of spin-angular momentum, it is, therefore, seeking a suitable qausiprobability distribution function for polarized light is susceptible for intensive investigations [33]. Furthermore, the optical field may possess higher-order polarization not characterized even by higher-order stokes parameters [25]. The macroscopic optical Schrodinger cat states [34], cat-like states [35] or entangled coherent states are typical instances of optical fields showcasing higher-order optical-polarization.

The entangled coherent states, a coinage due to Sanders, are introduced in proposing non-linear Mach-Zehnder interferometer as a device to generate superposition of a bimodal coherent states [36] which are applicable as continuous variable Bell-pair for variant applications in quantum information science [37] and quantum metrology [38]. Despite notable flurry of multi-faceted research activities in the last two decades experimental generation of entangled coherent states is recently reported [39]. The study of polarization and its description in quantum phase space of the optical field in Schrodinger cat, Cat-like and entangled coherent states has received little attention. Recently, one of the authors studied the optical-polarization of Schrodinger cat or cat-like states by generalizing the concept of usual theory of optical-polarization [40].

In the present investigation, the transiting operator (kernel operator) affecting the transition from Hilbert space to quantum phase space through Cahill-Glauber c(s)-correspondence rule is generalized (see Eq.10). This generalized transiting operator is utilized to evaluate polarization-HOWF ($2^{nd}$-order) for 'even' entangled coherent state (see Eq.14). It is, concomitantly, observed that for n being odd [41] HOWF suffers dependency on the intensities of light in bi-modes and it ends up with equivalent fate as Stokes theory in few photonic and intense regimes [42]. The nature of polarization-HOWF is analyzed on various probable characteristic polarization-parameters' values of 'even' entangled coherent states. The nature, immaculately, shows that individual peaks (for bi-modes) and inference-peak attain positions periodically in quantum phase space.

The paper is organized as follows: Section 2 introduces transiting operator for optical-polarization, in Section 3 we generalize Kernel-operator (transiting operator) which would facilitates transition of higher-order optical-polarized states from Hilbert space to quantum phase space (Planck space); and in Section 4 polarization of higher-order Wigner function (HOWF) for optical entangled coherent state in quantum phase space is evaluated and its dynamics is analyzed.

## 2. Transiting Operator for perfect optical-polarization

The dynamical state of a harmonic oscillator, specified by classical complex amplitude, α, may be described in quantum theory by analogous quantum complex amplitude (annihilation operator) and its density operator, $\hat{\rho}$ in Hilbert space. An alternative description of the same is provided by a quasi-distribution function in quantum phase space derived by following Cahill-Glauber C(s)-correspondence rule [9] with transiting operator (kernel operator), $\hat{t}$ (α, s) through the prescription, W(α, s) = Tr[ $\hat{\rho}$ $\hat{t}$ (α, s)], where W(α, s) is a parameterized phase-space distribution function. Notably, $\hat{t}$ (α, s) is the Fourier transform of the s- parameterized displacement operator, $\hat{D}$ (ξ, s), $\hat{t}$ (α, s) = $\pi^{-1}$ ∫ $\hat{D}$ (ξ, s) exp(αξ$^*$-α$^*$ξ) d$^2$ξ. Here $\hat{D}$ (ξ, s) bears simple relationship with unitary displacement operator, $\hat{D}$ (ξ), $\hat{D}$ (ξ, s) = $\hat{D}$ (ξ) exp(s|ξ|$^2$/2). Simple algebraic manipulation yields instructive expression for transiting operator, $\hat{t}$ (α, s) as

$$\hat{t}(\alpha, s) = \left(\frac{2}{1-s}\right) \hat{D}(\alpha) \left(\frac{s+1}{s-1}\right)^{\hat{a}^\dagger \hat{a}} \hat{D}^\dagger(\alpha) \tag{1a}$$

or, in separable form,

$$\hat{t}(\alpha, s) = \left(\frac{2}{1-s}\right) \exp\left(\frac{-2\ |\alpha|^2}{1-s}\right) \exp\left(\frac{2\alpha\hat{a}^\dagger}{1-s}\right) : \exp\left(\frac{-2\hat{a}^\dagger\hat{a}}{1-s}\right) : \exp\left(\frac{2\alpha^*\hat{a}}{1-s}\right), \tag{1b}$$

where : : represents normal ordering of operators therein and $\hat{a}$ ($\hat{a}^\dagger$) are usual bosonic annihilation (creation) operator satisfying standard commutation relation [$\hat{a}$, $\hat{a}^\dagger$]=1.

Optical-Polarization in Classical Optics is studied by superposition of two orthogonal harmonic oscillators with equal frequencies which emulates plane transverse orthogonal bi-modes of a beam of monochromatic light propagating along fixed direction (say z). The analytic signal of vector potential, $\vec{\mathcal{A}}$ of such optical field reads, in Classical Optics, as

$$\vec{\mathcal{A}} = [\hat{e}_x \alpha_x(t) + \hat{e}_y \alpha_y(t)] e^{ikz}, \tag{2}$$

or, in Quantum Optics, as

$$\vec{\mathcal{A}} = \left(\frac{2\pi}{\omega V}\right)^{1/2} [(\hat{e}_x \hat{a}_x(t) + \hat{e}_y \hat{a}_y(t)) e^{ikz} + \text{h.c.}], \tag{3}$$

where $\alpha_{x,y}(t)$ ($\hat{a}_{x,y}(t)$) are classical (quantum) complex amplitudes, $\alpha_{x,y}(t) = \alpha_{x,y} e^{-i\omega t}$ ($\hat{a}_{x,y}(t) = \hat{a}_{x,y} e^{-i\omega t}$); $\alpha_{x,y} = \alpha_{0x,0y} e^{i\phi_{x,y}}$, $\alpha_{0x,0y}$ are real amplitudes having extremely rapid and random spatio-temporal variation with angular frequency, ω; $\phi_{x,y}$ (phase parameters) have random values, $0 \leq \phi_{x,y} < 2\pi$, in linear polarization-basis $(\hat{e}_x, \hat{e}_y)$ for unpolarized light; h.c. stands for Hermitian conjugate, **k** $(= k\hat{e}_z)$ is propagation vector of magnitude k, and $\hat{e}_{x,y,z}$ are unit vectors along respective x-, y-, z-axes forming right handed triad. Singh and Prakash [27, 28] worked out a necessary criterion for perfect polarized light ($\hat{\rho}$) by introducing an optical-polarization operator which picks out the index of polarization, p, and furnishes the non-random characteristic optical-polarization parameters: 'ratio of real amplitudes', $\alpha_{0y}/\alpha_{0x}$ and 'difference in phase', $\phi_y - \phi_x$. The aforementioned criterion read as

$$\hat{a}_y \hat{\rho} = p\hat{a}_x \hat{\rho}, \qquad (4a)$$

where,

$$p = \frac{\alpha_y}{\alpha_x} = \tan\theta \, e^{i\Delta}. \qquad (4b)$$

Here $\alpha_{x,y}$ are parametrized through expressions, $\alpha_x = A_0 \cos\theta \exp(i\varphi_x)$ and $\alpha_y = A_0 \sin\theta \exp(i\varphi_y)$; $\Delta = \varphi_y - \varphi_x$, phase-difference and $A_0^2 \, (= |\alpha_x|^2 + |\alpha_y|^2)$ is intensity of light.

The state of perfect polarization of a bi-modal optical field specified by density operator, ρ in Hilbert space may be mapped to a probability distribution function in quantum phase space through a bi-modal global transiting operator. Since plane transverse orthogonal bi-modes of unpolarized light are, in general, statistically independent [21], the transiting operator $\widehat{T}(\alpha_x, \alpha_y, s)$ must, therefore, be expressed in the product form, $\widehat{T}(\alpha_x, \alpha_y, s) = \hat{t}(\alpha_x, s)\hat{t}(\alpha_y, s)$. Using Eq. (1) for each oscillators, one yields with an expression for global transiting operator for unpolarized light beam,

$$\widehat{T}(\alpha_x, \alpha_y, s) = \left(\frac{2}{1-s}\right)^2 \exp\left(\frac{-2(|\alpha_x|^2 + |\alpha_y|^2)}{1-s}\right) \exp\left(\frac{2(\alpha_x \hat{a}_x^\dagger + \alpha_y \hat{a}_y^\dagger)}{1-s}\right) : \exp\left(\frac{-2(\hat{n}_x + \hat{n}_y)}{1-s}\right) : \exp\left(\frac{2(\alpha_x^* \hat{a}_x + \alpha_y^* \hat{a}_y)}{1-s}\right), \quad (5)$$

where $\hat{n}_{x,y} \equiv \hat{a}_{x,y}^\dagger \hat{a}_{x,y}$. Insertion of Eq. (4) into Eq.(5) results in the expression of transiting operator, $\widehat{T}(\alpha_x, p, s)$ for the state of perfect polarized light,

$$\widehat{T}(\alpha_x, p, s) = \left(\frac{2}{1-s}\right)^2 \exp\left(\frac{-2(1+|p|^2)|\alpha_x|^2}{1-s}\right) \exp\left(\frac{2\alpha_x(\hat{a}_x^\dagger + p\hat{a}_y^\dagger)}{1-s}\right) : \exp\left(\frac{-2(\hat{n}_x + \hat{n}_y)}{1-s}\right) : \exp\left(\frac{2\alpha_x^*(\hat{a}_x + p^*\hat{a}_y)}{1-s}\right). \quad (6)$$

Recently, applying Eq.(6), polarization probability distribution function (Gaussian) for quadrature bi-modal coherent states of light is derived which is, precisely, the polar-phase space description [33]. A similar investigation for this and other states are also carried out by Klimov et.al. [33].

## 3. Transiting Operator for higher-order optical-polarization

Quite recently, the concept of higher-order optical polarization [40] is introduced by demanding non-random values for all multiple-powers of positive integer, n, say of the ratio of transverse orthogonal complex amplitudes except n = 1. The 'index of polarization' for light in higher ($n^{th}$)-order reads,

$$p^{(n)} \equiv (\alpha_y / \alpha_x)^n, \qquad (7a)$$

of which analogous quantum criterion is

$$(\hat{a}_y)^n \hat{\rho} = p^{(n)} (\hat{a}_x)^n \hat{\rho}. \qquad (7b)$$

Here $\hat{\rho}$ specifies the dynamical state of optical-field. The bi-modal optical-field prepared at Schrodinger cat state in at least one mode possesses typical instances of higher-order polarized light. Evidently, defining Eq.(7) for higher-order optical-polarization correspond to Eq.(4) for usual optical-polarization, if n attains unit value. Furthermore, polar decomposition of higher ($n^{th}$)-order 'index of polarization', $p^{(n)} = |p^{(n)}| \exp(i\Delta^{(n)})$, communicate us that while 'ratio of real amplitudes', $\alpha_{0y} / \alpha_{0x}$ has a non-random value, $|p^{(n)}|^{1/n}$, the 'difference in phases', $\varphi_y - \varphi_x$ may pick up equally probable n non-random values, $\frac{1}{n}(\Delta^{(n)} + 2m\pi)$ with m = 0, 1, 2,.....(n-1), in steps of ($2\pi/n$). It may be emphasized that the characteristic polarization parameters of optical-field in higher ($n^{th}$)-order polarization are numerically '1+n' because it has single 'ratio of real amplitudes' and 'n', 'difference in phases' compared to the only two non-random parameters: single 'ratio of real amplitudes' and single 'difference in phases' for usual optical-polarization ensuring its truly mono-modal nature. The possession of '1+n'-characteristic polarization-parameters for higher ($n^{th}$)-order polarized light offers in it weird features.

The consideration of higher ($n^{th}$)-order polarization, naturally, demands that Eq.(6) must be generalized to affect the transition from Hilbert space and facilitates the polarization description in quantum phase space of an optical field preserving higher-order polarization as

$$\hat{T}^{(n)} = \mathcal{J} \exp\left(\frac{2\alpha_x^n (\hat{a}_x^{\dagger n} + p^{(n)} \hat{a}_y^{\dagger n})}{1-s}\right) : \exp\left(\frac{-2 (\hat{n}_x^n + \hat{n}_y^n)}{1-s}\right) : \exp\left(\frac{2\alpha_x^{*n} (\hat{a}_x^n + p^{*(n)} \hat{a}_y^n)}{1-s}\right) \quad (10)$$

where $\mathcal{J} = \left(\frac{2}{1-s}\right)^2 \exp\left(\frac{-2(1+|p^{(n)}|^2)|\alpha_x|^{2n}}{1-s}\right)$ is being an intensity dependent constant; non-random complex parameters, $p^{(n)} \equiv \alpha_y^n / \alpha_x^n$ is 'index of polarization' in phase space. The higher ($n^{th}$)-order transiting operator, Eq.(10) will, now, be utilized to map the states of polarization of 'even' entangled coherent states from Hilbert space to quantum phase space for displaying the dynamics of polarization-($2^{nd}$-order) Wigner distribution function.

## 4. Higher ($2^{nd}$)-order polarization-Wigner function for Entangled coherent states

The 'even' entangled coherent states [43] may be written as

$$|\Psi\rangle = N_+(\beta, \gamma)(|\beta, \gamma\rangle + |-\beta, -\gamma\rangle), \quad (11)$$

where $N_+(\beta, \gamma) = [2(1 + e^{-\mathcal{J}_{sum}})]^{-1/2}$ with $\mathcal{J}_{sum} = 2(|\beta|^2 + |\gamma|^2)$ is the normalization constant. One may encounter paradoxial situation for few photonic regime ($|\gamma| = |\beta| \to 0$) and intense regime ($|\gamma| = |\beta| \to \infty$) [42]. The criterion of higher-order optical polarization, Eq.(7b) provides 'index of polarization',

$$p_{\mathcal{H}s}^{(n=2)} \equiv p_{\mathcal{H}s}^{(2)} = (\gamma/\beta)^2, \quad (12)$$

in the linear polarization basis of description ($\hat{e}_x, \hat{e}_y$) which deciphers non-random characteristic polarization parameters: $\frac{|\gamma|}{|\beta|} = \left|p_{\mathcal{H}s}^{(2)}\right|^{1/2}$ and $\varphi_\gamma - \varphi_\beta = \frac{1}{2}(2l\pi + \Delta_{\mathcal{H}s}^{(2)})$ with $l = 0, 1$, where $\beta = |\beta|\exp(i\varphi_\beta)$ and $\gamma = |\gamma|\exp(i\varphi_\gamma)$. Interestingly, the 'single' value of 'ratio of amplitudes', $\left|p_{\mathcal{H}s}^{(2)}\right|^{1/2}$ and two values of 'difference in phases' specifies intricately equally probable diagonally opposites states of polarization on Poincare sphere. With the help of Eqs.(10-11) one obtains polarization-Wigner distribution function (for $s = 0$) as

$$W^{(n)}(\alpha, s=0) = \text{Tr}[\hat{\rho}\,\hat{T}^{(n)}],$$

or, $W^{(n)}(\alpha_x^n, p^{(n)}, s = 0) = |N_+(\beta, \gamma)|^2 \mathcal{J}(\langle\beta, \gamma| + \langle-\beta, -\gamma|)\exp(2\alpha_x^n(\hat{a}_x^{\dagger n} + p^{(n)}\hat{a}_y^{\dagger n}))$

$$\times : \exp(-2(\hat{n}_x^n + \hat{n}_y^n)) : \exp(2\alpha_x^{*n}(\hat{a}_x^n + p^{*(n)}\hat{a}_y^n))(|\beta, \gamma\rangle + |-\beta, -\gamma\rangle). (13)$$

$2^{nd}$-order polarization-Wigner function is obtained by taking $n = 2$ [41],

$$W^{(2)}(\alpha_x, p^{(2)}) = \mathcal{J}' \exp\left[2\alpha_x^2 \beta^{*2}\left(1 + p^{(2)} p_{\mathcal{H}s}^{*(2)}\right) + \text{c.c.}\right],$$

where c. c. stands for complex conjugate and $\mathcal{J}' = 4\exp[-2(1 + |p^{(2)}|^2)|\alpha_x|^2 - 2(|\beta|^4 + |\gamma|^4)]$. Manipulations by polar decompositions of various complex quantities involved yields,

$$W^{(2)}(|\alpha_x|, \varphi_x, \Delta^{(2)}) = \mathcal{J}' \exp\left[4|\alpha_x|^2 |\beta|^2 (\cos\Phi + |p^{(2)}||p_{\mathcal{H}s}^{(2)}|\cos(\Phi + \Theta))\right], \quad (14)$$

where $\Phi = 2(\varphi_x - \varphi_\beta)$, $\Theta = \Delta^{(2)} - \Delta_{\mathcal{H}s}^{(2)}$ with $\alpha_x = |\alpha_x|\exp(i\varphi_x)$, $p^{(2)} = |p^{(2)}|\exp(i\Delta^{(2)})$, and $p_{\mathcal{H}s}^{(2)} = |p_{\mathcal{H}s}^{(2)}|\exp(i\Delta_{\mathcal{H}s}^{(2)})$. Due to exponentiated appearance of parameters $(|\alpha_x|, \varphi_x, \Delta^{(2)})$ in phase space the 2$^{\text{nd}}$-order polarization-Wigner function, Eq.(14) foisted to describe the 'states of polarization' of 'even' entangled coherent states never attains negative values, an uncanny feature of nonclassical states of light [44]. Furthermore, the 2$^{\text{nd}}$-order polarization-Wigner distribution function reveals pristine dependence on characteristic polarization parameters $(|p_{\mathcal{H}s}^{(2)}|, \Delta_{\mathcal{H}s}^{(2)})$ of optical field in entangled coherent state. For equally intense optical field in bi-modes, $|p_{\mathcal{H}s}^2| = 1$, and for unit Poincare sphere, $p^{(2)} = 1$, one may obtain simplified self-explanatory form of Eq.(14) as

$$W^{(2)}(|\alpha_x|, \varphi_x, \delta) = \mathcal{J}' \exp[8|\alpha_x|^2|\beta|^2 \cos(\delta - \delta_{\mathcal{H}s} - \pi k + \pi l) \cos(2(\varphi_x - \varphi_\beta) + \delta - \delta_{\mathcal{H}s} - \pi k + \pi l) \quad (15)$$

where $\delta = \frac{1}{2}(\Delta^{(2)} + 2m\pi)$ and $\delta_{\mathcal{H}s} = \frac{1}{2}(2l\pi + \Delta_{\mathcal{H}s}^{(2)})$ with m, l = 0, 1 and β being fixed arbitrary complex amplitude. Wigner function, Eq.(15), on variant values of parameters, δ and $\varphi_x$ keeping $|\alpha_x|$ and $|\beta|$ fixed ($|\alpha_x| = 0.8$ and $|\beta| = 0.7$) have been analyzed (see Figs.1). The nature of 2$^{\text{nd}}$-order polarization-Wigner function, clearly, demonstrates the appearance of interference peak between individual peaks of bi-modes. The exemplary feature of the dynamics is that the points of occurrence of interference-peak oscillates.

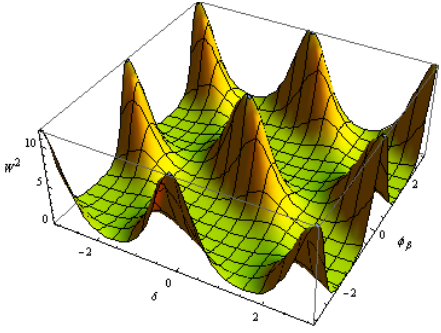

**Fig.1(a)** Wigner function with phase space parameters, $\delta$, $\varphi_x$, $|\alpha_x| = 0.8$) and characteristics polarization parameters: in phase space, having fixed values, $|\beta| = 0.7$, $\varphi_\beta = 0$, $\delta_{Hs} = 0$, m = 1; l = 0(for m and l see the text).

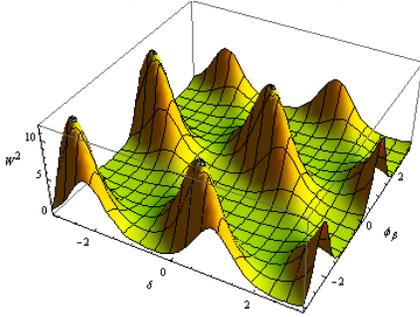

**Fig.1(b)** Wigner function with phase space parameters, $\delta$, $\varphi_x$, $|\alpha_x| = 0.8$) and characteristics polarization parameters: in phase space, having fixed values, $|\beta| = 0.7$, $\varphi_\beta = \pi/6$, $\delta_{Hs} = \pi/6$, m = 1; l = 0(for m and l see the text).

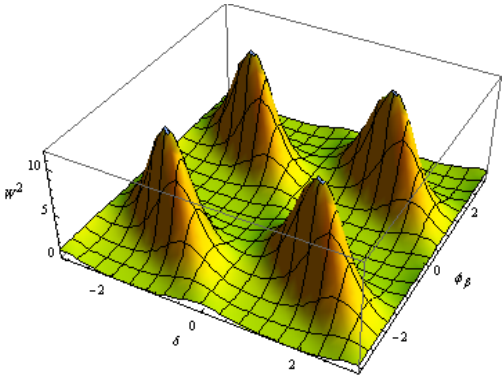

**Fig.1(c)** Wigner function with phase space parameters, $\delta$, $\varphi_x$, $|\alpha_x| = 0.8$) and characteristics polarization parameters: in phase space, having fixed values, $|\beta| = 0.7$, $\varphi_\beta = 3\pi/6$, $\delta_{Hs} = 3\pi/6$, m = 1; l = 0(for m and l see the text).

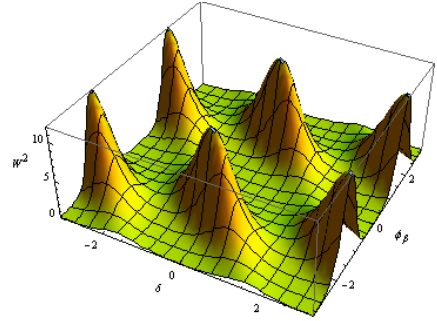

**Fig.1(d)** Wigner function with phase space parameters, $\delta$, $\varphi_x$, $|\alpha_x| = 0.8$) and characteristics polarization parameters: in phase space, having fixed values, $|\beta| = 0.7$, $\varphi_\beta = 3\pi/6$, $\delta_{Hs} = \pi$, m = 1; l = 0(for m and l see the text).

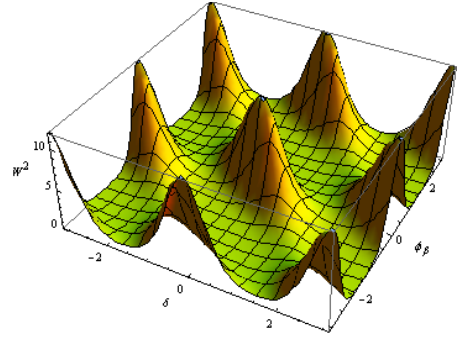

**Fig.1(e)** Wigner function with phase space parameters, $\delta$, $\varphi_x$, $|\alpha_x| = 0.8$) and characteristics polarization parameters: in phase space, having fixed values, $|\beta| = 0.7$, $\varphi_\beta = \pi$, $\delta_{Hs} = 0$, m = 1; l = 0(for m and l see the text).

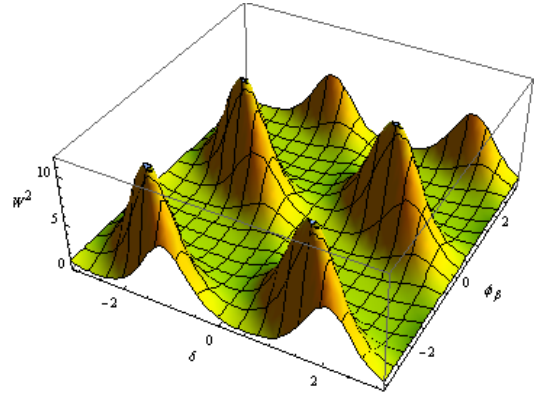

**Fig.1(f)** Wigner function with phase space parameters, $\delta$, $\varphi_x$, $|\alpha_x| = 0.8$) and characteristics polarization parameters: in phase space, having fixed values, $|\beta| = 0.7$, $\varphi_\beta = \pi/6$, $\delta_{Hs} = 3\pi/6$, m = 1; l = 1(for m and l see the text).

## Conclusion

Transiting operator (kernel operator) in Cahill-Glauber c(s)-correspondence rule is generalized to render higher ($n^{th}$)-order transiting operator. It is, then, used to map the higher ($2^{nd}$)-order polarization of entangled bi-modal coherent states into quantum phase space which furnishes, in turn, higher ($2^{nd}$)-order polarization-Wigner distribution function (HOWF). The nature of HOWF for 'even' entangled bi-modal coherent states have been analyzed graphically (see Figs.1). The presence of three peaks, in general, corresponds to 'two' individual modes and the higher peak can be attributed to interference between these bi-modes. The occurrence of interference-peaks is oscillating depending upon phase-space-parameters, $\delta$ and $\varphi_x$. The graphics, streakly, demonstrates that $2^{nd}$-order polarization-Wigner distribution function is of non-Gaussian nature possessing non-negative values. It may be noted that polarization-HOWF of 'odd' entangled bi-modal coherent states can, similarly, be derived and discussed.

**Acknowledgement**: The authors gratefully acknowledge the critical yet inspirational discussions with Professors H. Prakash and R. Prakash, University of Allahabad, Allahabad, India. One of the authors (RSS) acknowledges invoking comments imparted by Prof. Vijay A. Singh, HBCSE, Tata Institute of Fundamental Research, Mumbai, India and Prof. Surendra P. Singh, Department of Physics, University of Arkansas, Fayetteville, U. S. A.

Electronic addresses: [#]yesora27@gmail.com, [#]nisaly06@rediffmail.com, [*]gyankg@gmail.com